\documentclass[
    ,final            
  ]
  {aipproc}

\layoutstyle{8x11single}

\usepackage{natbib}

\usepackage{epsfig}

\begin{document}


\font\ksmss=cmss10     scaled \magstephalf
\font\kvsmss=cmss8     scaled \magstephalf

\newcommand{\cge}    {{$_ >\atop{^\sim}$}}
\newcommand{\cle}    {{$_ <\atop{^\sim}$}}
\newcommand{\n}	     {\noindent}%
\newcommand{\si}	 {\smallskip\indent}%
\newcommand{\sn}	 {\smallskip\noindent}%
\newcommand{\mi}	 {\medskip\indent}%
\newcommand{\mn}	 {\medskip\noindent}%
\newcommand{\bi}	 {\bigskip\indent}%
\newcommand{\bn}	 {\bigskip\noindent}%
\newcommand{\cl}     {\centerline}%
\newcommand{\ve}     {\vfill\eject}
\newcommand{\eg}	 {\mbox{e.g.,} }%
\newcommand{\cf}	 {\mbox{c.f.,} }%
\newcommand\etal     {{et\thinspace al.}}     
\newcommand{\ie}	 {\mbox{i.e.} }%
\newcommand{\aka}	 {\mbox{a.k.a.\ }}%
\newcommand{\bul}    {$\bullet$\ }
\newcommand\Bj       {{$B_{J}$}}
\newcommand{\HST}	 {\emph{HST}}%
\newcommand{\Ha}	 {\emph{H-$\alpha$}}%
\newcommand\JAB      {{$J_{AB}$}}
\newcommand\Lstar    {{$L^{*}$}}
\newcommand\mum      {{$\mu$m}}
\newcommand\Msun     {{\ $M_{\odot}$}}
\newcommand\Siv      {{$S_{1.4}$} }
\newcommand\muJy     {{$\mu$Jy} }
\newcommand\mjj      {{$\mu$Jy}}
\newcommand\zform    {{$z_{form}$}}
\newcommand\zAB      {{$z_{AB}$} }
\newcommand{\Astro}	 {\emph{Astro}}%
\newcommand\sspt     {{$\buildrel{s}           \over .$}}
\newcommand\degpt    {{$\buildrel{\circ}       \over .$}}
\newcommand\adegpt   {{$\buildrel{\circ}       \over .$}}
\newcommand\arcmpt   {{$\buildrel{\prime}      \over .$}}
\newcommand\arcspt   {{$\buildrel{\prime\prime}\over .$}}
\newcommand\magpt    {{$\buildrel{m}           \over .$}}
\newcommand\adeg     {{\hskip .10em {}^{\circ}\hskip -.40em . \hskip.25em} }
\newcommand\amin     {{\hskip .10em {}'       \hskip -.30em . \hskip.20em} }
\newcommand\asec     {{\hskip .10em {}''      \hskip -.49em . \hskip.29em} }
\newcommand\A{{{\kvsmss A}{\hspace*{-2.0mm}\vspace*{-0.0mm}$^{\scriptsize o}$}\vspace*{+0.0mm}}}
\newcommand\degsq    {{$deg^{2}$}}
\newcommand\sqdeg    {{$deg^{-2}$}} 
\newcommand{\kmsMpc}	{\mbox{km s$^{-1}$ Mpc$^{-1}$}}%
\newcommand{\magarc}	{{$\mbox{mag arcsec}^{-2}$}}%
\newcommand{\SB}	 {{$\mu_{\scriptsize\sl F300W}$}}%
\newcommand{\re}	 {{$r_e$}}%
\newcommand{\rhl}	 {{$r_{hl}$} }%
\newcommand{\HI}	 {\mbox{H\,{\sc i}}}%
\newcommand{\HII}	 {\mbox{H\,{\sc ii}}}%
\newcommand\Kmorph   {{$K_{morph}$} }
\newcommand{\UIT}	 {\emph{UIT}}%
\newcommand\zmed     {{$z_{med}$}}
\newcommand{\tsim}	 {{\sim}}%
\newcommand{\lsim}	 {\makebox[15pt]
           {\mbox{{\raisebox{-3pt}{$\sim$}}{\llap{\raisebox{2pt}{$<$}}}}}}%
\newcommand{\gsim}	 {\makebox[15pt]
           {\mbox{{\raisebox{-3pt}{$\sim$}}{\llap{\raisebox{2pt}{$>$}}}}}}%
\newcommand\degree   {{\ifmmode^\circ\else$^\circ$\fi}} 

\newcommand\AAP      {A\&A, }
\newcommand\aap      {A\&A}
\newcommand\AAL      {A\&AL, }
\newcommand\aal      {A\&AL}
\newcommand\AAPS     {A\&ApS, }
\newcommand\aaps     {A\&ApS}
\newcommand\AAS      {A\&AS, }
\newcommand\aas      {A\&AS}
\newcommand\ACTAA    {Acta~A, }
\newcommand\AJ       {AJ, }
\newcommand\aj       {AJ}
\newcommand\AN       {AN, }
\newcommand\APLET    {Ap. Let., }
\newcommand\APJ      {ApJ, }
\newcommand\apj      {ApJ}
\newcommand\APJL     {ApJL, } 
\newcommand\apjl     {ApJL} 
\newcommand\APL      {ApJ, }  
\newcommand\APJS     {ApJS, }
\newcommand\apjs     {ApJS}
\newcommand\APS      {ApJS, }
\newcommand\APSS     {Ap\&SS, }
\newcommand\apss     {Ap\&SS}
\newcommand\ARAA     {ARA\&A, }
\newcommand\araa     {ARA\&A}
\newcommand\AUSJP    {Australian J. Phys., }
\newcommand\AZH      {AZh, }
\newcommand\BAAS     {BAAS, }
\newcommand\baas     {BAAS}
\newcommand\FCP      {Fund. Cosmic Phys., }
\newcommand\HOA      {Highlights Astr., }
\newcommand\IAU      {IAU Symp., }
\newcommand\ITR      {Internal Technical Report of the Netherlands Foundation for 
                      Radio Astronomy, No.} 
\newcommand\JOSAMA   {J. Opt. Soc. Am. A, }
\newcommand\JRASC    {JRASC, }
\newcommand\MEMRAS   {MmRAS, }
\newcommand\MNRAS    {MNRAS, }
\newcommand\mnras    {MNRAS}
\newcommand\NAT      {{\it Nature}, }
\newcommand\nat      {{\it Nature}}
\newcommand\NPS      {Nat. Phys. Sc., }
\newcommand\PASA     {PASA, }
\newcommand\pasa     {PASA}
\newcommand\PASJ     {PASJ, }
\newcommand\pasj     {PASJ}
\newcommand\PASP     {PASP, }
\newcommand\pasp     {PASP}
\newcommand\PHD      {Ph.D. thesis, }
\newcommand\PHYSCR   {Physica Scripta, }
\newcommand\QJRAS    {QJRAS, }
\newcommand\REP      {Reports on Astronomy, IAU Transactions, }
\newcommand\REPPPHY  {Rep. Prog. Phys., }
\newcommand\REVMP    {Rev. Mod. Phys., }
\newcommand\SCAM     {Sc. Am., }
\newcommand\SCI      {Science, }
\newcommand\SOVAST   {Sov. Astr., }
\newcommand\SPIE     {SPIE, }
\newcommand\ST       {S\&T, }
\newcommand\VIS      {Vistas in Astronomy, }
\newcommand\VLA      {VLA Test Memorandum, No.}

\newlength{\txw}
\setlength{\txw}{\textwidth}
\newlength{\txh}
\setlength{\txh}{\textheight}



\title{
``GiGa'': the Billion Galaxy HI Survey --- Tracing 
Galaxy Assembly from Reionization to the Present}


\author{R. A. Windhorst, S. H. Cohen, N. P. Hathi, R. A. Jansen, R. E. Ryan}{
address={School of Earth \& Space Exploration, Arizona State University,
Box 871404, Tempe, AZ 85287-1404, USA;\ \ \ Email:\ Rogier.Windhorst@asu.edu}
}

\begin{abstract} 

In this paper, we review the Billion Galaxy Survey that will be carried out at
radio--optical wavelengths to micro--nanoJansky levels with the telescopes of
the next decades. These are the Low-Frequency Array, the Square Kilometer
Array and the Large Synoptic Survey Telescope as survey telescopes, and the
Thirty Meter class Telescopes for high spectral resolution+AO, and the James
Webb Space Telescope (JWST) for high spatial resolution near--mid IR follow-up.
With these facilities, we will be addressing fundamental questions like how
galaxies assemble with super-massive black-holes inside from the epoch of First
Light until the present, how these objects started and finished the reionization
of the universe, and how the processes of star-formation, stellar evolution,
and metal enrichment of the IGM proceeded over cosmic time. 

We also summarize the high-resolution science that has been done thus far on
high redshift galaxies with the Hubble Space Telescope (HST). Faint galaxies
have steadily decreasing sizes at fainter fluxes and higher redshifts,
reflecting the hierarchical formation of galaxies over cosmic time. HST has
imaged this process in great structural detail to z\cle 6. We show that
ultradeep radio-optical surveys may slowly approach the natural confusion
limit, where objects start to unavoidably overlap because of their own sizes, 
which only SKA can remedy with HI redshifts for individual sub-clumps. Finally,
we summarize how the 6.5 meter James Webb Space Telescope (JWST) will measure
first light, reionization, and galaxy assembly in the near--mid-IR.


\end{abstract}

\maketitle


\section{Introduction}

For this review paper, I was asked to review the ``Billion Galaxy Survey'' that
will be carried out at radio--optical wavelengths to micro--nanoJansky levels
with the telescopes of the next decades. These facilities are, for instance,
the Low-Frequency Array (LOFAR, R\"ottgering et al. 2005), the Square
Kilometer Array (SKA; Schilizzi\ etal 2004), and the Large Synoptic Survey
Telescope (LSST; Tyson 2007) --- which will be used as survey telescopes at
radio optical wavelengths --- and the Thirty Meter Telescopes (TMT; \eg
Nelson \etal 2006) --- including \eg the Giant Segmented Mirror Telescope,
Giant Magellan Telescope, plus the EU Extremely Large Telescope ---which will
be used for high spectral resolution+ adaptive optics follow-up, as well as the
James Webb Space Telescope (Mather and Stockman, 2000), which will provide high
spatial resolution near--mid-IR imaging and low-resolution spectroscopy. With a
combination of these facilities available in the next decade, we will be
addressing fundamental questions like: 

\n \bul (1) How do HI clouds at z\cge 6 assemble over cosmic time into the
giant spiral and elliptical galaxies seen today? 

\n \bul (2) How and why did the (dwarf dominated) galaxy luminosity function
(LF) and mass function evolve with epoch? 

\n \bul (3) In the context of the galaxy formation--AGN paradigm, how did
supermassive black hole (SMBH) growth keep up with the process of galaxy
assembly? 

\n \bul (4) How does the central accretion disk feed the SMBH, and how are
radio jets and lobes in radio galaxies and quasars produced as a result? 

\n \bul (5) How did AGN feedback control the bright-end evolution of the galaxy
LF, and how did SN feedback shape the faint-end of the LF from z\cge 6 to z=0? 

These are some of critical science drivers for radio and optical telescopes of
the next decade. Before we give a preview of possible answers to these
questions, we will first briefly consider how the Hubble Space Telescope (HST)
has revolutionized the topic of galaxy assembly in the last decade. One of the
remarkable discoveries by HST was that the numerous faint blue galaxies are in
majority late-type (Abraham \etal\ 1996, Glazebrook \etal\ 1995, Driver \etal\
1995) and small (Odewahn \etal\ 1996, Pascarelle \etal\ 1996) star-forming
objects. These are the building blocks of the giant galaxies seen today. By
measuring their distribution over rest-frame type versus redshift, HST has
shown that galaxies of all Hubble types formed over a wide range of cosmic
time, but with a notable transition around redshifts 


\ve 

\noindent\begin{minipage}[b]{0.74\txw}
\psfig{file=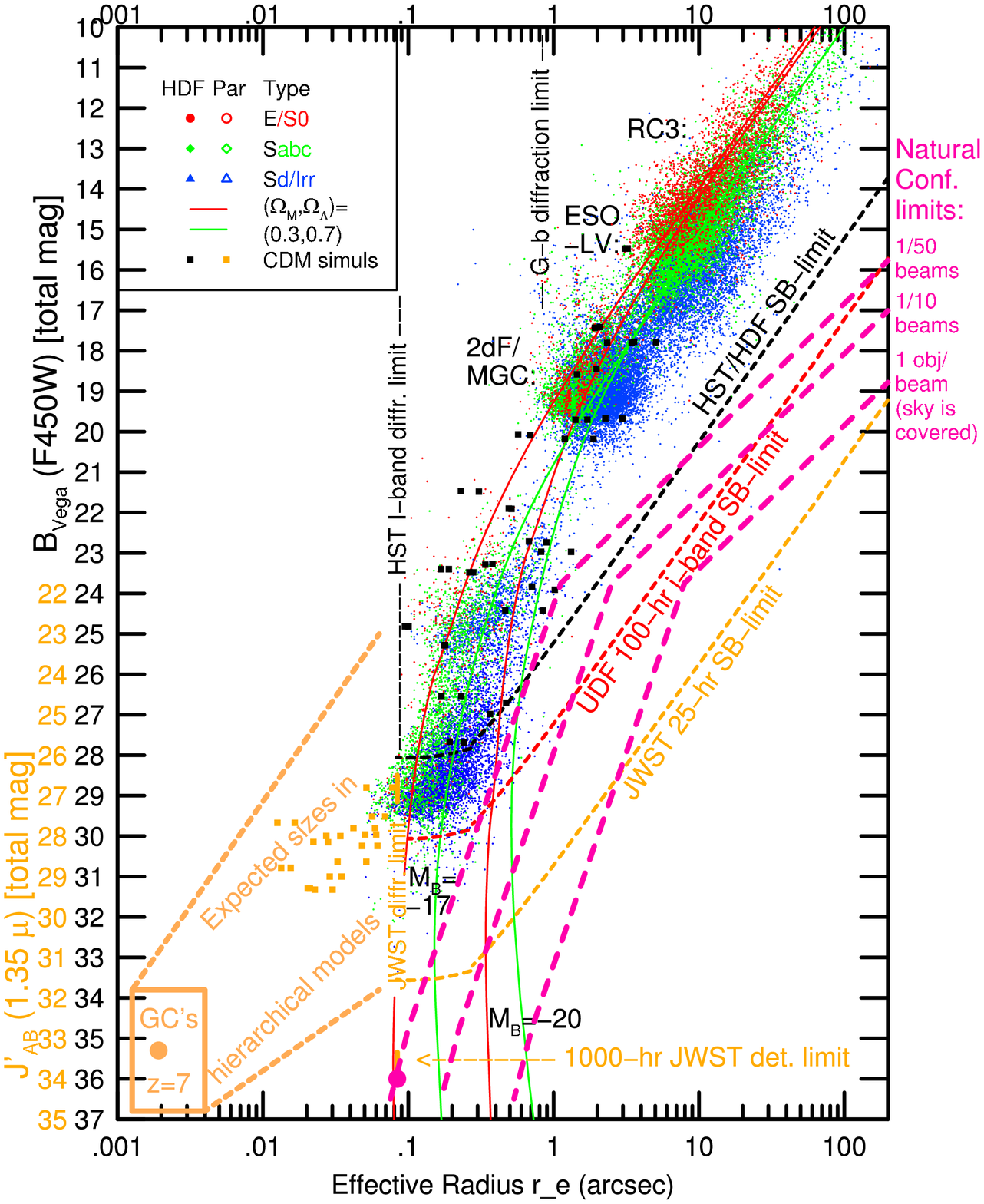,angle=0,width=0.73\txw,height=0.575\txh} 
\end{minipage}
\begin{minipage}[b]{0.2525\txw}
\noindent{\footnotesize
{\bf Fig. 1} \ Galaxy sizes vs. B$_{Vega}$ or \JAB-mag from the RC3 to the HUDF
limit. Short dashed lines indicate survey limits for the HDF (black), HUDF
(red), and JWST (orange): the point-source sensitivity is horizontal and the
SB-sensitivity has slope=+5 mag/dex. Broken long-dashed pink lines indicate the
natural confusion limit, below which objects begin to overlap due to their own
sizes. Red and green lines indicate the expectations at faint fluxes of the
{\it non-evolving} median size for RC3 elliptical and spiral galaxies,
respectively (Odewahn \etal\ 1996). Orange and black squares indicate
hierarchical size simulations (Kawata \etal 2003). Note that most galaxies at
\JAB\cge 28 mag are expected to be smaller than the HST and JWST diffraction
limits (\ie r$_{hl}$\cle 0\arcspt 1).\\

\n \\

}
\end{minipage}


\bn z$\simeq$0.5--1.0 (\eg Driver \etal\ 1998, Elmegreen \etal\ 2007). This was
done through HST programs like the Medium-Deep Survey (Griffiths \etal\ 1994),
GOODS (Giavalisco \etal\ 2004), GEMS (Rix \etal\ 2005), and COSMOS (Scoville
\etal\ 2007). Subgalactic units rapidly merged from the end of reionization to
grow bigger units at lower redshifts (Pascarelle \etal\ 1996). Merger products
start to settle as galaxies with giant bulges or large disks around redshifts
z$\simeq$1 (Lilly \etal\ 1998, 2007). These evolved mostly passively since
then, resulting in giant galaxies today, possibly because the epoch-dependent
merger rate was tempered at z\cle 1 by the extra expansion induced by $\Lambda$
(Cohen \etal\ 2003, Ryan \etal\ 2007). To avoid caveats from the morphological
K-correction (Giavalisco \etal\ 1996, Windhorst \etal\ 2002), galaxy structural
classification needs to done at rest-frame wavelengths longwards of the Balmer
break at high redshifts (Taylor-Mager \etal\ 2007). JWST will make such studies
possible with 0\arcspt 1--0\arcspt 2 FWHM resolution at observed near--IR
wavelengths (1--5 \mum), corresponding to the restframe optical--near-IR at the
median redshift of faint galaxies (\zmed$\simeq$1--2; Mobasher \etal\ 2007). 

\section{Radio \& Optical Sizes at the Faintest Fluxes: Natural Confusion?}

The HST/ACS GOODS survey (Ferguson \etal\ 2004) showed that the median sizes of
faint galaxies decline steadily towards higher redshifts, despite the
$\Theta$--z relation that minimizes at z$\simeq$1.65 in WMAP $\Lambda$CDM
cosmology. While surface brightness (SB) and other selection effects in these
studies are significant, this work suggests evidence for intrinsic size
evolution of faint galaxies, where galaxy half-light radii \rhl evolve with
redshift as:\ r$_{\rm hl}$(z)\,$\propto$\,r$_{hl}$(0)$\cdot$(1+z)$^{\rm -s}$\ 
with s\,$\simeq$\,1. This reflects the hierarchical formation of galaxies, 
where sub-galactic clumps and smaller galaxies merge over time to form the
larger/massive galaxies that we see today (\eg Navarro, Frenk, \& White 1996).

\ve 

\n{\begin{minipage}[b]{0.42\txw}
 \psfig{file=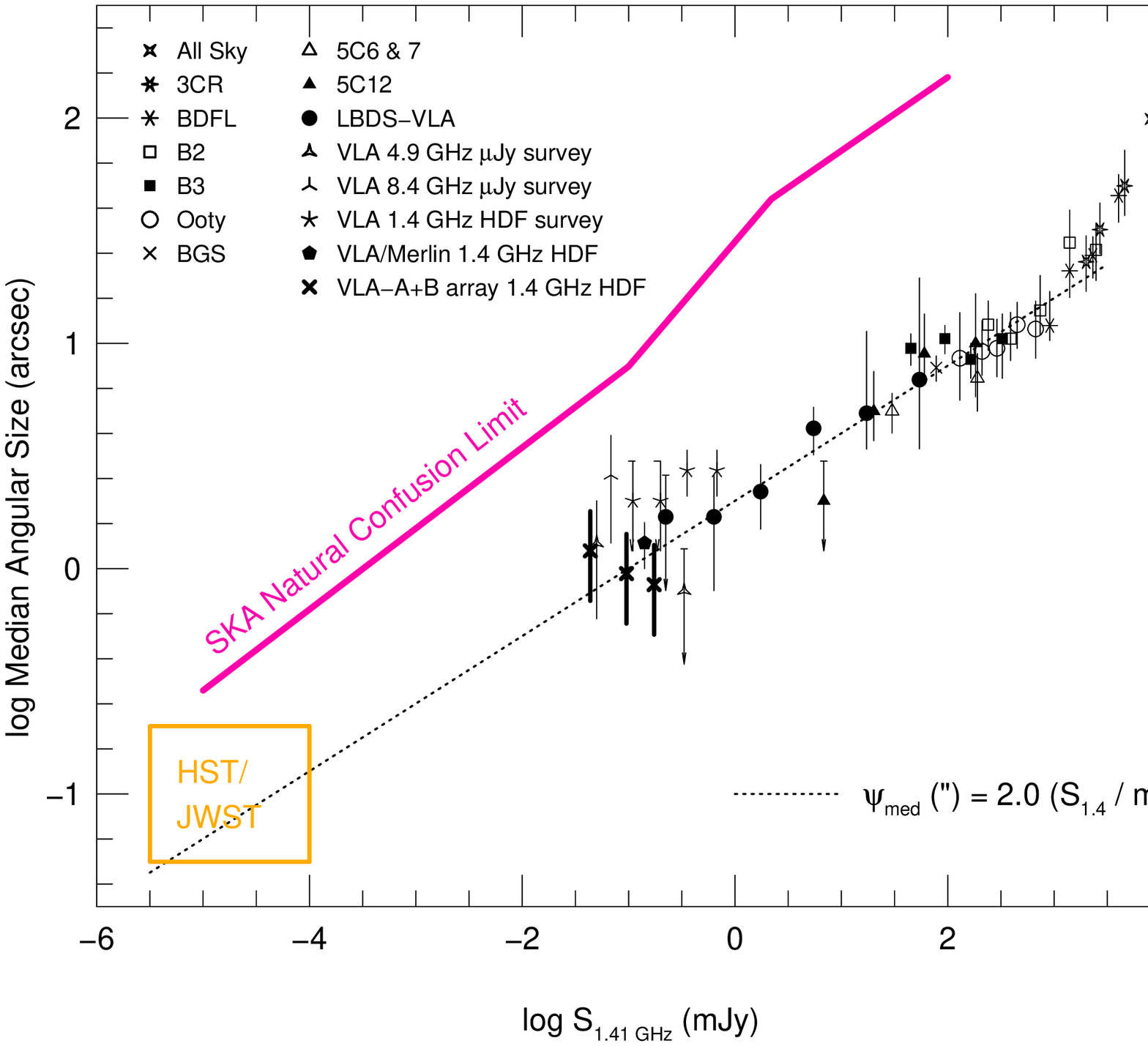,angle=0,width=0.42\txw} 
\end{minipage}\hspace*{2cm}
\begin{minipage}[b]{0.45\txw}
{\footnotesize
{\bf Fig. 2} \ Median 1.4 GHz radio source size vs. \Siv flux density, as
summarized by Windhorst \etal (1993, 2003). The dotted line indicates the
median extrapolated to the rest-frame UV--optical disk sizes seen by HST at
AB\cle 30 mag, or expected for JWST at AB\cle 32 mag. The pink line indicates
the natural confusion limit (see Fig. 1), above which radio sources unavoidably
start to overlap in ultradeep surveys, requiring HI information to disentangle
them.}\vspace*{3.5cm}\null
\end{minipage}


\bn

\bn

\n\cl{
\includegraphics[angle=-0.0,width=0.490\txw]{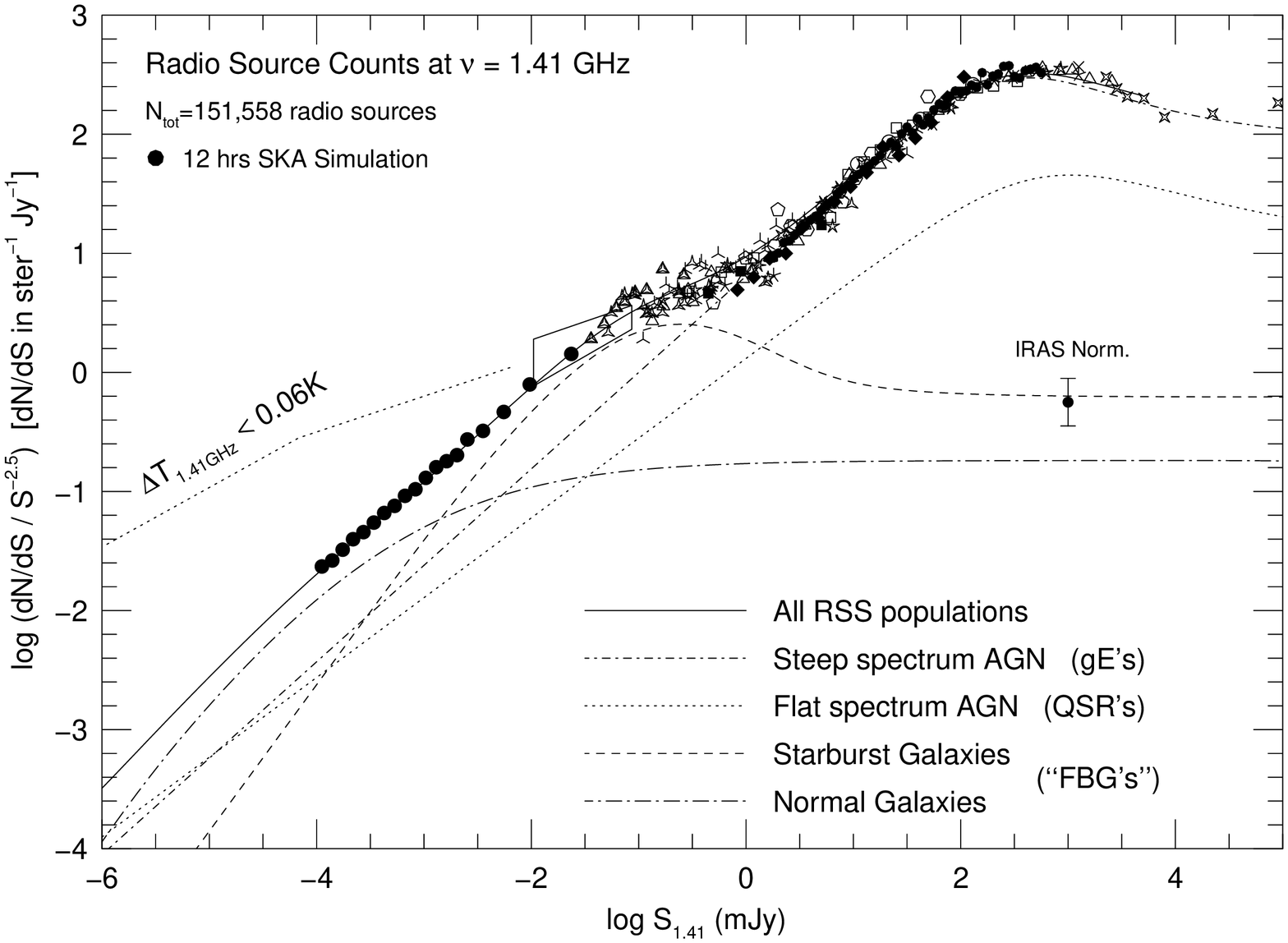}\ \ \
\includegraphics[angle=-0.0,width=0.485\txw]{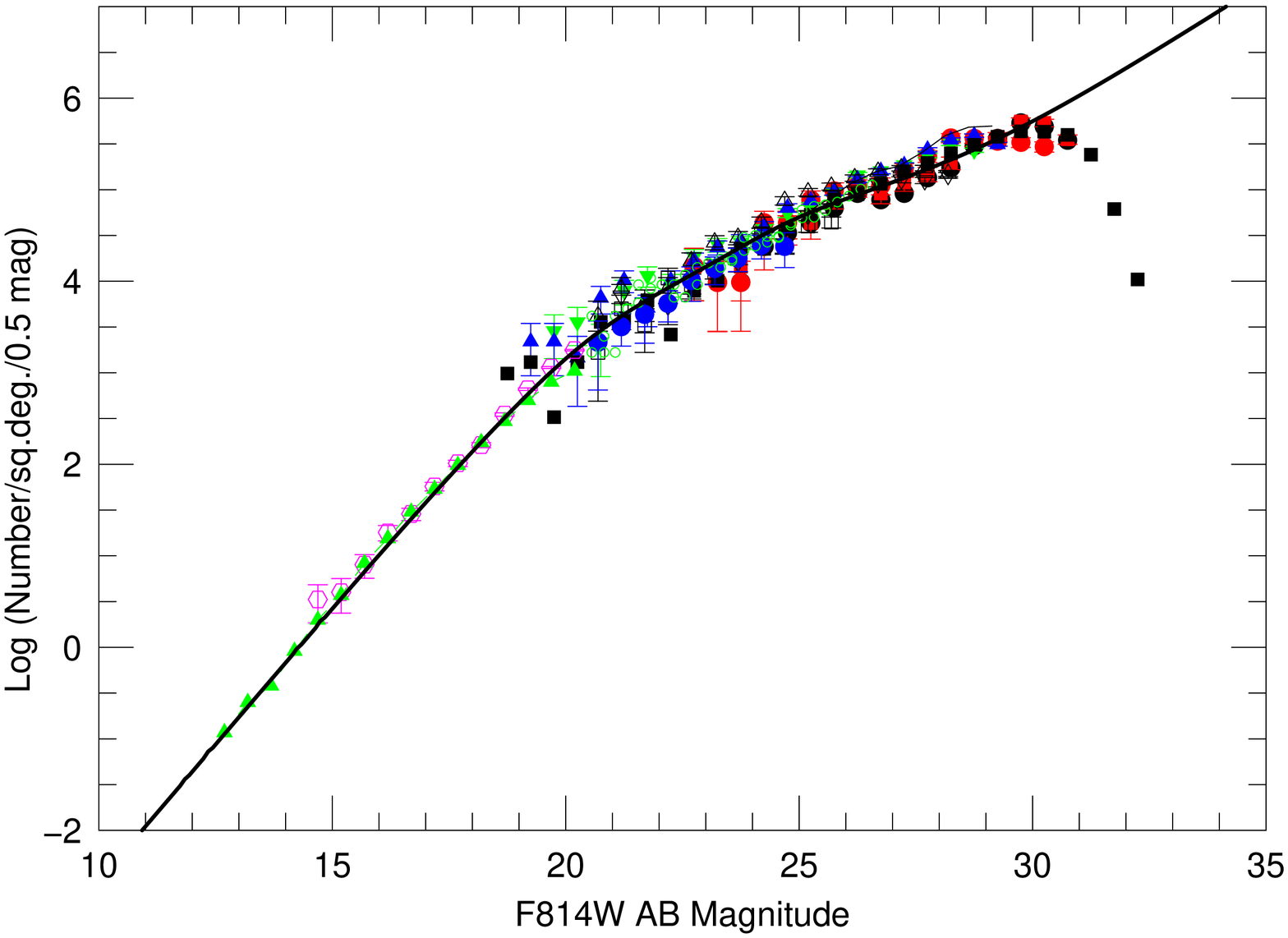}
}

\mn{\footnotesize
{\bf Fig. 3a} \ Differential 1.4 GHz radio source counts normalized to
Euclidean, as summarized by Windhorst \etal (1993, 2003) with models from
Hopkins \etal\ (2000) below 10 \muJy.
{\bf Fig. 3b} \ Optical HUDF galaxy counts in i'-band. To the completeness limit
of AB\cle 30 mag (4 nJy at 1\mum), there are over 10$^6$ galaxies \degsq, so
that \cge 10$^3$ needs to be surveyed to get one billion galaxies.}

\bn The HST/ACS Hubble UltraDeep Field (HUDF; Beckwith \etal\ (2006) showed that
high redshift galaxies are intrinsically very small, with typical sizes of
$r_{\rm hl}$$\simeq$ 0\arcspt 12 or 0.7--0.9 kpc at z$\simeq$4--6. A
combination of ground-based and HST surveys shows that the apparent galaxy
sizes decline steadily from the RC3 to the HUDF limits (Fig. 1 here; Odewahn
\etal\ 1996; Cohen \etal\ 2003, Windhorst \etal\ 2006). At the bright end, this
is due to the survey SB-limits, which have a slope of +5 mag/dex in Fig. 1. At
the faint end, ironically, this appears {\it not only} to be due to
SB-selection effects (cosmological (1+z)$^{4}$ SB-dimming), since for \Bj\cge
23 mag the samples do {\it not} bunch up against the survey SB-limits. Instead
it occurs because: (a) their hierarchical formation and size evolution; (b) at
\JAB\cge 26 mag, one samples the faint end of the luminosity function (LF) at
z$_{\rm med}$\cge 2--3, resulting in intrinsically smaller galaxies (see \eg
Fig. 4b; Yan \& Windhorst 2004b); and (c) the increasing inability to properly
deblend faint galaxies at fainter fluxes. 

This leads ultradeep surveys to slowly approach the ``natural'' confusion
limit, where a fraction of the objects unavoidably overlaps with neighbors due
to their finite \emph{object size} (Fig. 1), rather than the finite
instrumental resolution, which causes the \emph{instrumental} confusion limit.
Most galaxies at \JAB\cge 28 mag are likely unresolved point sources at
r$_{hl}$\cle 0\arcspt 1 FWHM, as suggested by hierarchical size simulations in
Fig. 1 (Kawata \etal\ 2002). This is, amongst others, why such in the
UV--optical--near-IR objects are best imaged from space, which provides the
best point-source and SB-sensitivity. The fact that many faint objects remain
unresolved at the HST diffraction limit effectively reduces the (1+z)$^{4}$
SB-dimming to a (1+z)$^{2}$ flux-dimming (with potentially an intermediate case
for partially resolved objects, or linear objects that are resolved in only one
direction), mitigating the incompleteness of faint galaxy samples. The trick in
ultradeep surveys is therefore to show that this argument has not become
circular, and that larger galaxies at high redshift are not missed. Other
aspects that compound these issues are size-overestimation due to object
confusion, size-bias due to the sky-background and due to image noise. 


Fig. 2 shows the median angular size of 1.4 GHz radio sources vs. \Siv flux
from 100 Jansky to 30 \muJy (\eg Windhorst \etal\ 2003). The SKA radio source
sizes at 10--100 nJy are estimated from the HST galaxy size distribution to
AB=30 mag in Fig. 1, where the object density reaches \cge 10$^6$
objects/deg$^2$ (see Fig. 3a). Fig. 1--2 clearly show that SKA must anticipate
the small \cle 0\arcspt 3 radio sizes of faint star-forming galaxies at all
redshifts z\cge 1. As in Fig. 1, the purple line in Fig. 2 indicates the
natural confusion limit due to intrinsic radio source sizes. Above this line,
radio sources unavoidably start to overlap statistically, and would
significantly impact catalog reliability and completeness, and --- unless other
measures are taken (such as disentangling the sources with very deep HI-imaging
and redshifts, see below) --- eventually fundamentally limit our ability to
survey the sky to arbitrarily deep limits. The HI disk sizes of nearby dwarf
galaxies are similar to their synchrotron radio continuum-sizes and their
optical sizes (Deeg \etal 1997). Hence, Fig. 2 suggests that SKA needs to have
baselines at $\sim$0\arcspt 3 FWHM to match the HI and radio continuum sizes 
that we expect from the galaxy disk sizes in the deepest HST images (Fig. 1). 

\section{The Billion Galaxy Survey at Radio and Optical Wavelengths}

To produce a billion galaxy survey, let us now consider what such a survey 
requires. Fig. 3a shows the HUDF galaxy counts to AB\cle 30 mag ($\simeq$4 nJy at
1\mum), and Fig. 3b shows the normalized differential 1.4 GHz radio source
counts expected for SKA to \Siv$\simeq$1--10 nJy. From the HUDF galaxy counts, we
expect \cge 2$\times$10$^6$ galaxies/deg$^2$ to AB=31.5 mag (1 nJy at
$\simeq$1\mum), which is the deepest level currently reachable with HST, and
which will be routinely obtained with JWST (see below). A similar flux level of
1--10 nJy at GHz frequencies is the main goal for the SKA. Hopkins \etal\
(2000) show a simulated 12-hour SKA 1.4 GHz image with 1 deg$^2$ FOV that has
FWHM$\simeq$0\arcspt 1 and reaches a flux level of 100 nJy (5-$\sigma$).
Similarly, a 1200 hr SKA integration would reach $\sim$10 nJy. Fig. 3a shows
that the normalized differential SKA source counts at nJy levels will be
dominated by blue, mostly point-like or disk-shaped radio sources. These will
likely primarily reside in star-forming galaxies at redshifts z$\simeq$1--2,
where the cosmological volume maximizes in WMAP cosmology, while a minority
will be --- possibly more extended --- radio sources that reside in
earlier-type galaxies which host weak AGN (see the models in Fig. 3a). 

To obtain a billion galaxies at either radio or optical wavelengths, one must
thus at least survey $\sim$10$^3$ deg$^2$, or a significant fraction of the
high-latitude sky. {\it Since neither HST, nor JWST, nor the TMT's will have a
large enough FOV to do so}, this can only be done with SKA at radio
wavelengths, and with the LSST at optical wavelengths, provided that the LSST
will survey 2$\pi$ ster or 2$\times$$10^4$\sqdeg to its anticipated all-sky
flux limit of AB$\simeq$27 mag. Fig. 1--3 show that at the radio flux level of
\Siv$\simeq$1--10 nJy, one expects 1 object in every box 2\arcspt
5$\times$2\arcspt 5. This means that one must always carry out SKA-surveys at
nJy levels with sufficient spatial resolution to avoid object confusion, and
that SKA must have coverage at significantly long baselines that yield 
FWHM$\simeq$0\arcspt 3. For HST and JWST, this number is FWHM\cle 0\arcspt 08,
as discussed in Fig. 1. For SKA, this means practically that one must always
obtain many HI line channels, so that one can disentangle overlapping continuum
sources in redshift space, and find all the enclosed HI. Nature simply does not
leave us an alternative.

In summary, with a dedicated Billion Galaxy Survey, SKA will measure how
galaxies of all types assembled their HI and turned it into stars over a wide
range of cosmic time: from z\cle 6 to z=0. The SKA HI and radio continuum sizes
of 10$^7$--10$^9$\Msun\ starforming objects are small enough that with \cle
0\arcspt 3 FWHM resolution, SKA will: (1) properly separate these objects from
their neighbors without major source confusion; and (2) not resolve them in
majority, somewhat mitigating the effects from cosmological SB-dimming. 
Further details on the future of HI surveys and their other main science goals 
are given by Drs. M. Haynes, J. Lazio, and S. Meyers (this Volume). 

\section{First Light, Reionization \& Galaxy Assembly with JWST} 

The James Webb Space Telescope (JWST) is designed as a deployable 6.5 meter
segmented IR telescope for imaging and spectroscopy from 0.6 \mum\ to 28 \mum.
After its planned 2013 launch (Mather \& Stockman 2000), JWST will be
automatically deployed and inserted into an L2 halo orbit. It has a nested
array of sun-shields to keep its temperature at \cle 40 K, allowing faint
imaging to AB\cle 31.5 mag ($\simeq$1 nJy) and spectroscopy to AB\cle 29 mag in
the near--mid-IR. 

{\bf First Light:}\ The WMAP polarization results imply that the Dark Ages
which started at recombination (z$\simeq$1089) lasted until the First Light
objects started shining at z\cle 20, and that the universe was first reionized
at redshifts at

\ve 

\null\vspace*{-0.5cm}
\n\makebox[\txw]{
\psfig{file=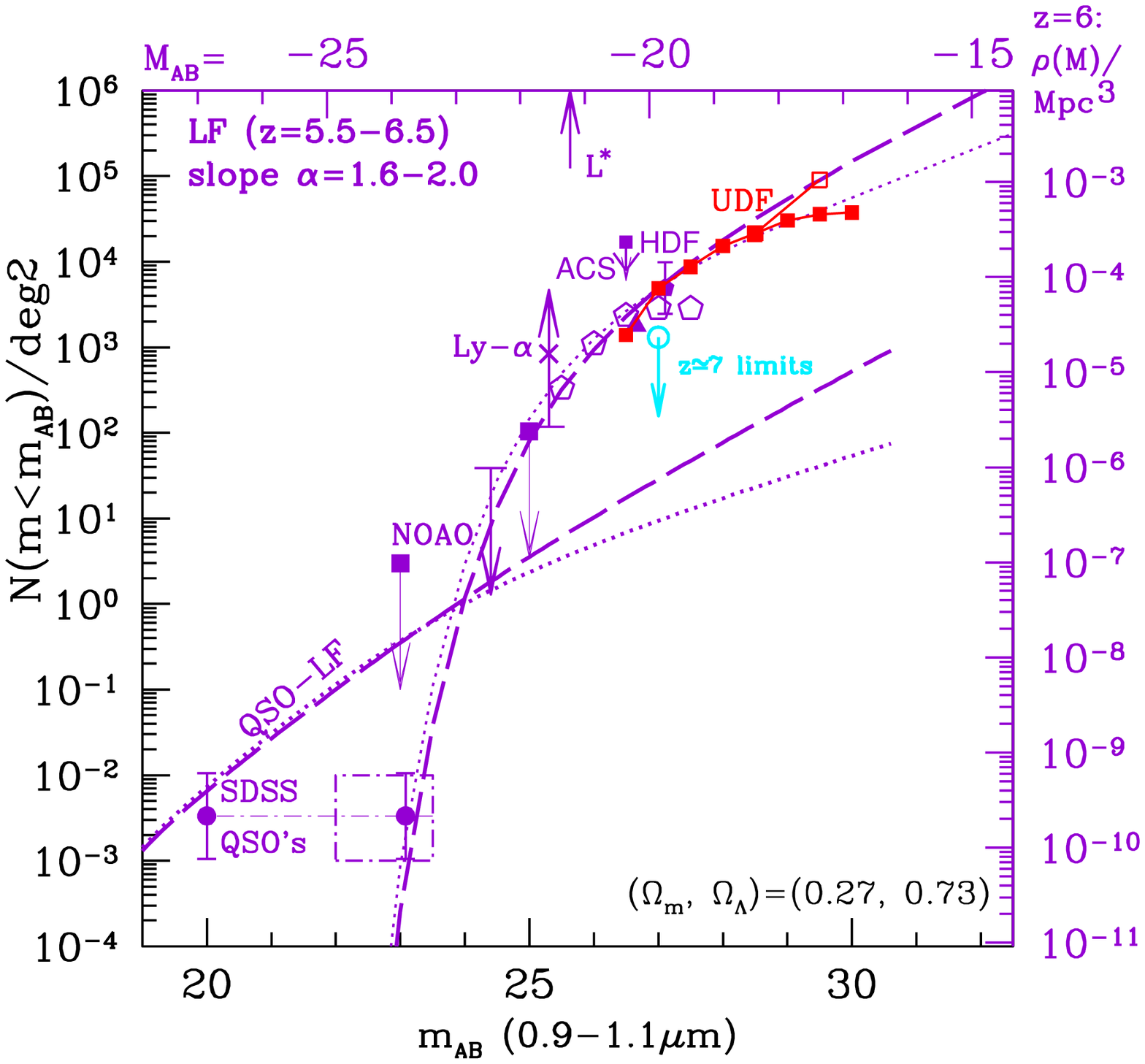,width=0.50\txw}\ \ 
\psfig{file=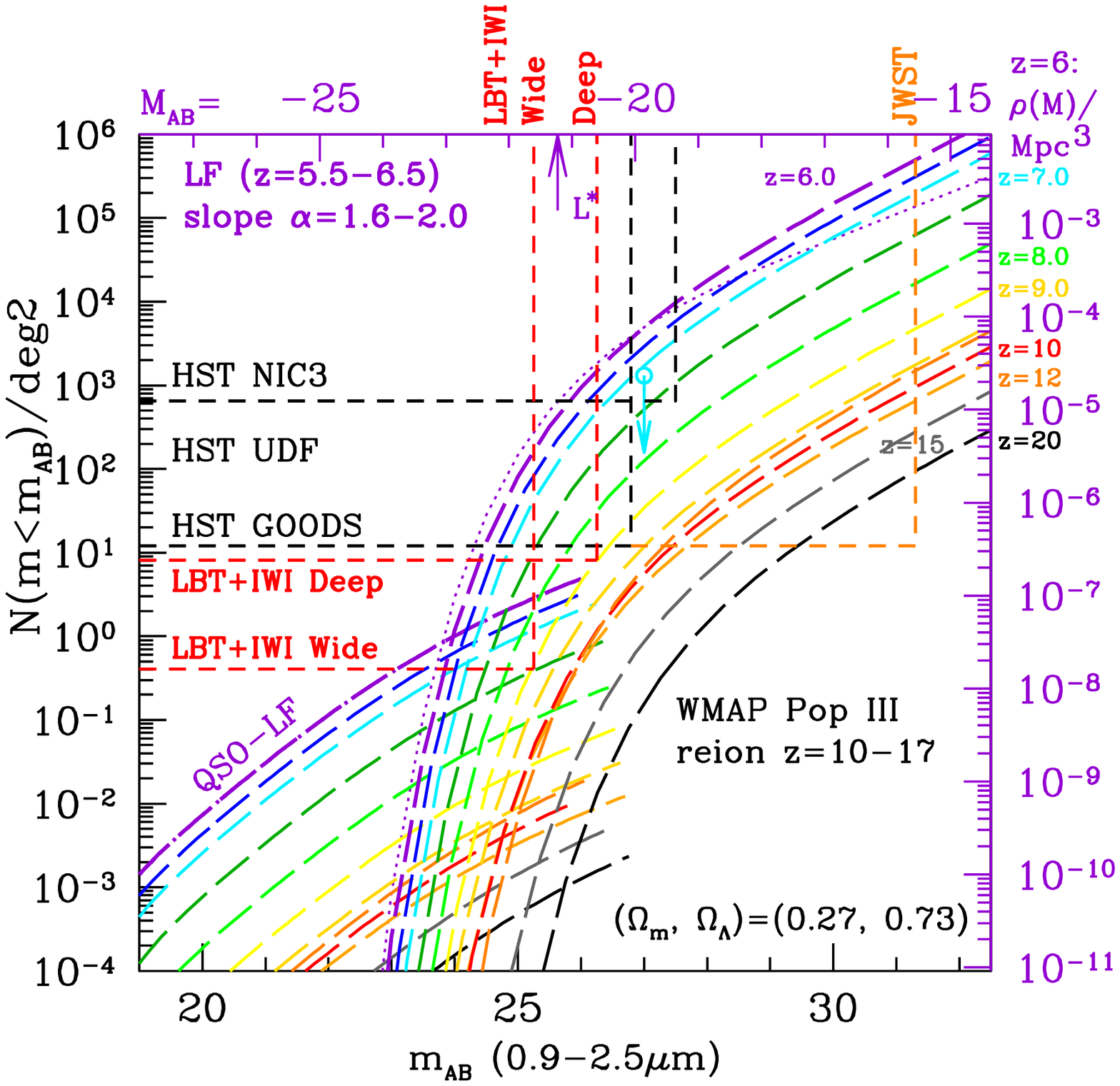,width=0.50\txw}
}
\null
{\footnotesize
\noindent{\bf Fig.~4a} \ Integral luminosity function (LF) of z$\simeq$6
objects, plotted as surface density vs. AB-mag. The z$\simeq$6 LF may be very
steep, with faint-end Schechter slope $\vert$$\alpha$$\vert$$\simeq$1.8--1.9
(Yan \& Windhorst 2004b). Dwarf galaxies and not quasars therefore likely
completed the reionization epoch at z$\simeq$6 (Yan \etal\ 2004a). This is
what JWST will observe in detail to AB$\simeq$31.5~mag (1~nJy). \ 
{\bf Fig.~4b} \ Possible extrapolation of the LF of Fig.~4a for z\cge 7, which
is not yet constrained by data. Successive colors show redshift shells 0.5 in
$\Delta$z apart from z=6, 6.5, ..., 10, and also for z=12, 15, 20. The HST/ACS
has detected objects at z\cle 6.5, but its discovery space
A$\cdot$$\Omega$$\cdot$$\Delta$log($\lambda$) is limited to z\cle 6.5. NICMOS
similarly is limited to z\cle 8 (Bouwens \etal\ 2004, Yan \& Windhorst 2004b).
JWST can trace the entire reionization epoch from First Light at z$\simeq$20 to
the end of reionization at z$\simeq$6.}

\bn z$\simeq$9--13, if reionization happened instantaneous, or as early as
z$\simeq$11--20, if the reionization process was drawn out (Dunkley \etal\
2008, Spergel \etal\ 2007). The epoch of First Light is thought to have
started with Population III stars of 200-300\Msun\ at z\cge 10--20 (Bromm
\etal\ 2003). Groupings of Pop III stars and possibly their extremely luminous
supernovae should be visible to JWST at z$\simeq$10--20 (Gardner \etal\ 2006).
This is why JWST needs NIRCam at 0.6--5 \mum\ and MIRI at 5--28 \mum. The
First Light epoch and its embedded Pop III reionizing sources may have been
followed by a delayed epoch of Pop II star-formation, since Pop III supernovae
may have heated the IGM enough that it could not cool and form the IMF of the
first Pop II stars until z\cle 8--10 (Cen 2003). The IMF of Pop II stars may
have formed in dwarf galaxies with masses of 10$^6$--10$^9$\Msun\ with a
gradual onset between z$\simeq$9 and z$\simeq$6. The reionization history may
have been more complex and/or heterogeneous, with some Pop II stars forming in
sites of sufficient density immediately following their Pop III predecessors at
z\cge 10. 

HST/ACS can detect objects at z\cle 6.5, but its discovery space
A$\cdot$$\Omega$$\cdot$$\Delta$log($\lambda$) cannot trace the entire
reionization epoch. HST/NICMOS similarly is limited to z\cle 8 and provides
limited statistics. HST/WFC3 can explore the redshift range z$\simeq$7--8 with
a wider FOV than NICMOS. Fig. 4b shows that with proper survey strategy (area
{\it and} depth), JWST can trace the LF throughout the entire reionization
epoch, starting with the first star-forming objects in the First Light epoch at
z\cle 20, to the first star-forming dwarf galaxies at the end of the
reionization epoch at z$\simeq$6. To observe the LF of First Light
star-clusters and subsequent dwarf-galaxy formation may require JWST to survey
GOODS-sized areas to AB$\simeq$31.5 mag ($\simeq$ 1 nJy at 10-$\sigma$; See
Fig. 4b), using 7 filters for reliable photometric redshifts, since objects
with AB\cge 29 mag will be too faint for spectroscopy.


\n {\bf Reionization:}\ The HUDF data showed that the LF of z$\simeq$6 objects
is potentially very steep (Bouwens \etal\ 2006, Yan \& Windhorst 2004b), with a
faint-end Schechter slope $\vert$$\alpha$$\vert$$\simeq$1.8--1.9 after
correcting for sample incompleteness (Fig. 4a). Deep HST/ACS grism spectra
confirmed that 85--93\% of HUDF i-band dropouts to \zAB\cle 27 mag are at
z$\simeq$6 (Malhotra \etal\ 2005). The steep faint-end slope of the z$\simeq$6
LF implies that dwarf galaxies may have collectively provided enough UV-photons
to complete reionization at z$\simeq$6 (Yan \& Windhorst 2004a). This assumes
that the Lyman continuum escape fraction at z$\simeq$6 is as large as observed
for Lyman Break Galaxies at z$\simeq$3 (Steidel \etal\ 1999), which is
reasonable --- although not proven --- given the expected lower dust content in
dwarf galaxies at z$\simeq$6. Hence, dwarf galaxies, and not quasars, likely
completed the reionization epoch at z$\simeq$6. The Pop II stars in dwarf
galaxies therefore cannot have started shining {\it pervasively} much before
z$\simeq$7--8, or no neutral H-I would be seen in the foreground of z\cge 6
quasars (Fan \etal\ 2003), and so dwarf galaxies may have ramped up their
formation fairly quickly from z$\simeq$9 to z$\simeq$6.

\ve 

\n\cl{
\includegraphics[width=0.500\txw,angle=0,clip=]{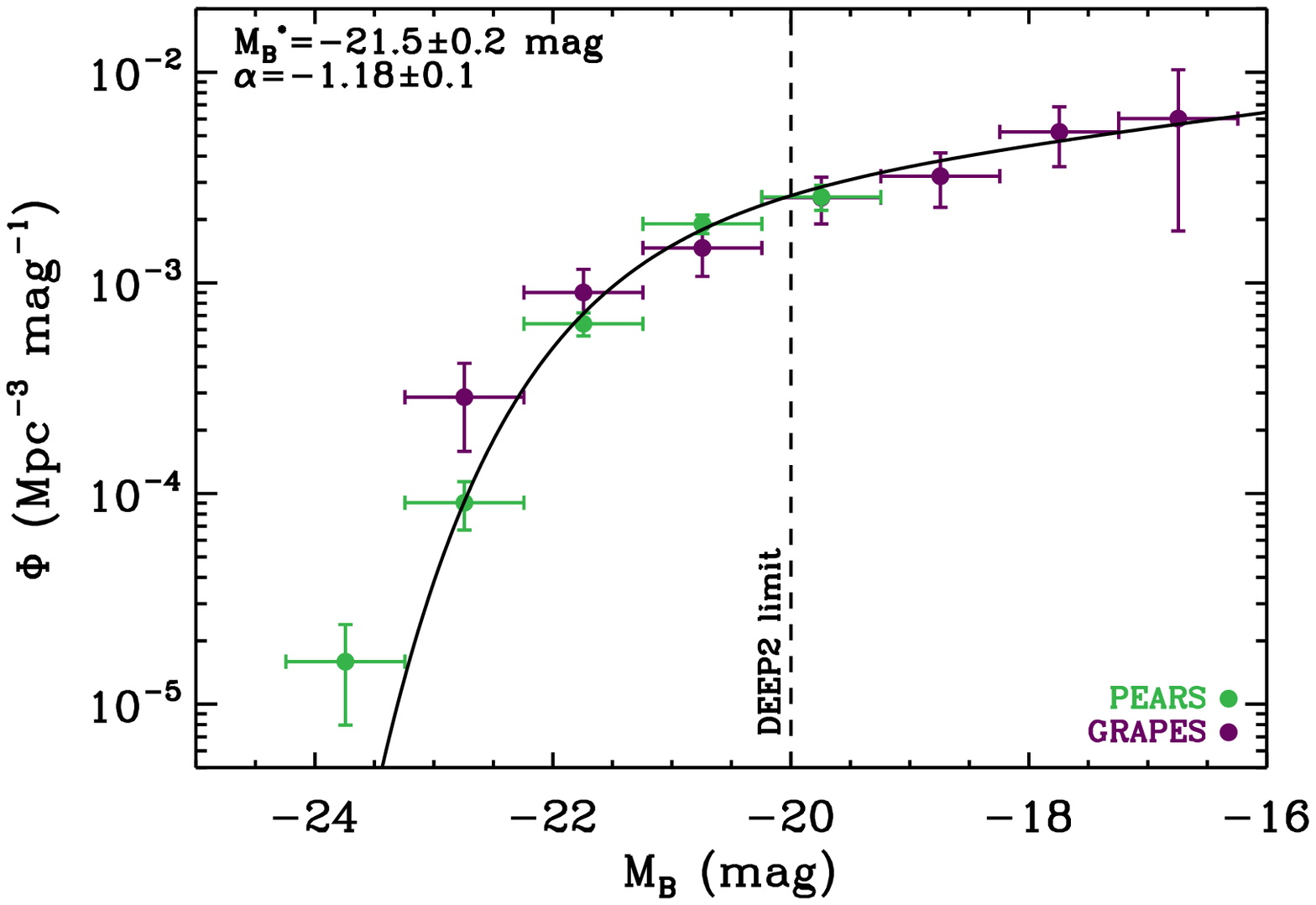}
\includegraphics[width=0.485\txw,angle=0,clip=]{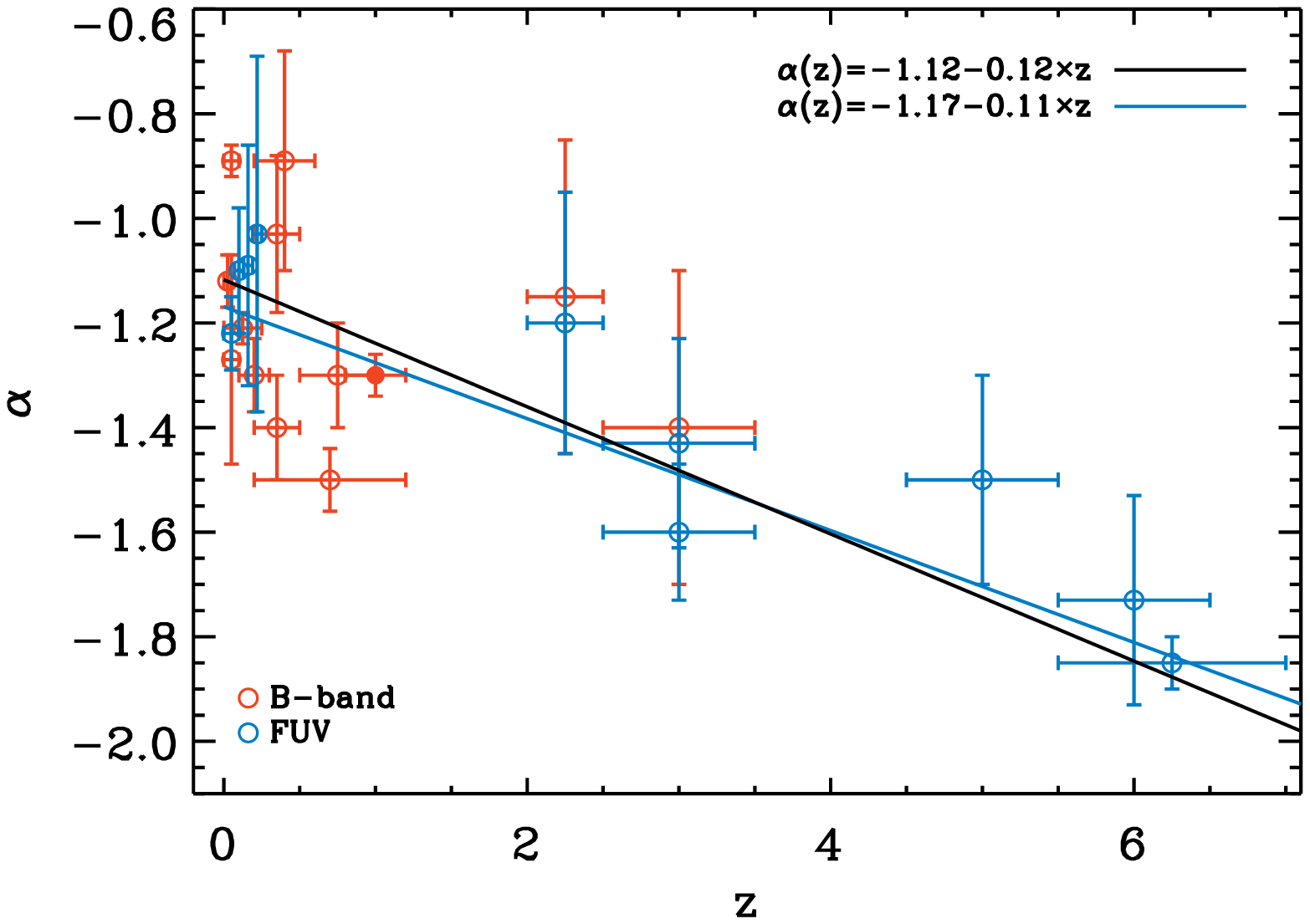}
}
\null
{\footnotesize
\noindent{\bf Fig.~5a} \ HST ACS grism LF from the HST GRAPES and PEARS surveys
to AB\cle 27 mag (Ryan \etal\ 2007). \ 
{\bf Fig.~5b} \ Faint-end LF-slope or Schechter-$\alpha$ evolution (Ryan \etal\
2007) from z$\simeq$0 to z$\simeq$6 for a variety of surveys, that reach several
magnitudes below \Lstar in the rest-frame FUV or rest-frame optical. 
}

\bn JWST surveys are designed to provide \cge 10$^4$ objects at z$\simeq$7 and
100's of objects in the epoch of First Light and at the start of reionization
(Fig. 4b). SKA will be critical to image the bright-end/higher mass end of
their HI-mass function at high redshifts, and delineate the process of galaxy
assembly from HI clouds over cosmic time. 


\n {\bf Galaxy Assembly:}\ JWST will measure how galaxies of all types formed
over a wide range of cosmic time, by accurately measuring their distribution
over rest-frame optical type and structure as a function of redshift or cosmic
epoch. HST/ACS has made significant progress at z$\simeq$6, surveying very
large areas (GOODS, GEMS, COSMOS), or using very long integrations (HUDF,
Beckwith \etal\ 2006). Quantitative galaxy decomposition (Odewahn \etal\ 2002)
can measure galaxy morphology and structure, and the the presence and
evolution of bars, rings, spiral arms, and other structural features at higher
redshifts (\eg Jogee \etal 2004). Such techniques will allow JWST to measure
the detailed history of galaxy assembly in the epoch z$\simeq$1--3, when most
of today's giant galaxies were made. JWST will be able to do this out to
z$\simeq$10-15 at least (see Windhorst \etal\ 2006), hence enabling to
quantitatively trace galaxy assembly. 



The red boundaries in Fig. 4b indicate part of the galaxy and QSO LF that a
ground-based 8m class telescope with a wide-field IR-camera can explore to
z\cle 9 and AB\cle 25 mag. A ground-based {\it wide-field} near-IR survey to
AB\cle 25--26 mag can sample L$>>$\Lstar galaxies at z\cle 9, which is an
essential ingredient to study the co-evolution of supermassive black-holes and
proto-bulges for z\cle 9, and an essential complement to the JWST First Light
studies. The next generation of wide-field near-IR cameras on ground-based 
8--30 m class telescopes can do such surveys over many \degsq\ to
AB$\simeq$25--27 mag, complementing JWST, which will survey GOODS-sized areas
to AB\cle 31.5 mag (Fig. 4b). 

HST has begun to measure the faint-end LF-slope evolution of the galaxy
luminosity function. This Schechter $\alpha$-evolution is a fundamental issue,
like the local low-mass end of the initial mass function (IMF). With accurate
ACS grism+broad-band redshifts to AB\cle 27 mag, Ryan \etal\ (2007) measured
the faint-end LF-slope at z\cge 1, and outlined the faint-end LF-slope
evolution from z$\simeq$0 to z$\simeq$6. As modeled by Khochfar \etal (2007),
this provides new physical constraints to hierarchical formation theories.
Star-formation and SN-feedback processes can further produce different
faint-end slope-evolution (Khochfar \etal\ 2007). JWST will provide fainter
spectra (AB\cle 29) and spectro-photometric redshifts to much higher z (\cle
15--20), and will trace the faint-end LF-slope evolution for z\cle 12, hence
constraining hierarchical formation theories.

Similarly, SKA, and perhaps LOFAR, will trace low-mass-end of the HI
mass-function of dwarf galaxies and its $\alpha$-slope evolution for z\cle
3--6. This will allow us to measure environmental impact on faint-end LF-slope
$\alpha$ directly. As known for nearby galaxy samples (Fig. 5a), significant
scatter is likely due to different clustering environments or cosmic variance.
We expect convergence to a slope $\vert$$\alpha$$\vert$$\simeq$2 predicted by
hierarchical models at z$>$6, before the process of SN feedback starts
destroying dwarf galaxies, or before the process of AGN feedback starts
producing major outflows and AGN-driven winds in massive galaxies. With a
combination of JWST and very deep SKA images, it may be possible to constrain
onset of Pop III SNe epoch at z\cge 8--10, and the onset to the Type II \& Type
Ia SN-epochs at z\cle 6, and thereby constraining the major physical processes
that regulate star-formation as function of cosmic time.

\section{Conclusions}

HST has led the study of galaxy assembly, showing that galaxies form
hierarchically through repeated mergers with sizes growing steadily over time
as r$_{\rm hl}$(z) $\propto$ r$_{\rm hl}$(0)$\cdot$(1+z)$^{\rm -s}$ and
s$\simeq$ 1. The Hubble sequence thus gradually emerges at z\cle 1--2, when
the epoch-dependent merger rate starts to wind down. The global onset of Pop
II-star dominated dwarf galaxies ended the process of reionization at
z$\simeq$6. High resolution rest-frame UV--optical imaging of high redshift
galaxies is best done from space, because faint galaxies are small (\rhl\cle
0\arcspt 15), while the ground-based sky is too bright and the PSF not stable
enough to obtain good high-resolution images at faint fluxes (AB\cge 27 mag). 
JWST will extend these studies into the epoch of reionization and First Light,
and trace galaxy SED's in the restframe-optical for z\cle 20. 

Such surveys will slowly approach the natural confusion limit (Fig. 1), where
objects at the faintest (nJy) fluxes start to unavoidably blend with their
neighbors, not because of instrumental resolution, but because of their own
intrinsic sizes. SKA will be critical to help disentangle the non-negligible
fraction of such objects at $\mu$Jy--nJy levels, and provide unique HI-redshifts
for each component. For this, it will need to have beam-sizes as small as
0\arcspt 3 FWHM (see Fig. 1). In order to get the next radio facilities it
wants, the science community will need to unify behind current \& future radio
facilities, such as SKA and LOFAR --- building on proto-types that are currently
being build (see this Vol.) --- and define the essential synergy of SKA with
other future facilities, such as the JWST, LSST, and the TMT's.



\begin{theacknowledgments}
This work was supported by HST grants from STScI, which is operated by AURA
for NASA under contract NAS 5-26555, and by NASA JWST grant NAG 5-12460. Other
JWST studies are at:\ www.asu.edu/clas/hst/www/jwst/ . 
\end{theacknowledgments}





\bibliographystyle{aipproc} 



\hspace*{-0.0cm}\vspace*{-0.0cm}
\n 

\ve 

\baselineskip=12pt

\end{document}